\documentclass[11pt,a4paper]{article}

\advance\voffset by -2.0cm \advance\hoffset by -1.25cm
\usepackage{amsmath}
\usepackage{amssymb}
\usepackage{amsthm}
\usepackage{amsfonts}

\numberwithin{equation}{section}

\theoremstyle{definition}

\newcommand{\RR}{\mathbb{R}} 
\newcommand{\ZZ}{\mathbb{Z}} 

\newcommand{\be}{\begin{equation}}
\newcommand{\ee}{\end{equation}}

\def\1{\frak 1}
\def\2{\frak 2}
\def\3{\frak 3}

\newlength{\oldcolsep}

\begin{document}

\title{Exponentiating Higgs}
\author{Marco Matone}\date{}

\maketitle

\begin{center} Dipartimento di Fisica e Astronomia ``G. Galilei'' \\
 Istituto
Nazionale di Fisica Nucleare \\
Universit\`a di Padova, Via Marzolo, 8-35131 Padova,
Italy\end{center}

\bigskip

\begin{abstract}

We consider two related formulations for mass generation in the $U(1)$ Higgs-Kibble model and in the Standard Model (SM).
In the first model there are no scalar self-interactions and, in the case of the SM, the formulation is related to the normal
subgroup of $G=SU(3)\times SU(2)\times U(1)$, generated by $(e^{2\pi i/3}I,-I,e^{\pi i/3})\in G$, that acts trivially on all
the fields of the SM. The key step of our construction is to relax the non-negative definiteness condition for the Higgs field
due to the polar decomposition. This solves several stringent problems, that we will shortly review, both in the perturbative
and non-perturbative formulations.
We will show that the usual polar decomposition of the complex scalar doublet $\Phi$ should be done with $U\in SU(2)/\ZZ_2\simeq SO(3)$,
where $\ZZ_2$ is the group generated by $-I$, and with the Higgs field $\phi\in\RR$ rather than $\phi\in\RR_{\geq0}$. As a byproduct, the
investigation shows how Elitzur's theorem may be avoided in the usual formulation of the SM.
It follows that the simplest lagrangian density for the Higgs mechanism has the standard kinetic term in addition to the mass term,
with the right sign, and to a linear term in $\phi$. The other model concerns the scalar theories with normal ordered exponential interactions.
The remarkable property of these theories is that for $D>2$ the purely scalar sector corresponds to a free theory.

\end{abstract}

\newpage

\section{Introduction}

The Higgs mechanism \cite{Englert:1964et}-\cite{Kibble:1967sv} is a basic step in the formulation of the Standard Model (SM)
\cite{Weinberg:1967tq,Salam:1968rm}.
This has been confirmed by the spectacular experimental results at LHC \cite{Aad:2012tfa,Chatrchyan:2012ufa}. Despite this, there are still some open questions.
The most important one is that the vev of the Higgs field is evaluated at the classical level. On the other other hand,
there are models with non-trivial minima for the classical potential, with no order parameter. The point
is that Spontaneous Symmetry Breaking (SSB) is a strictly non-perturbative phenomenon, concerning infinitely many degrees of freedom. As such, even radiative corrections
to $\langle\phi\rangle$ should be considered with particular attention.
In this respect, one should also recall that, against the evidence coming from the perturbative expansion, there is strong evidence that $\lambda\phi^4$ is a free theory,
that is, the renormalized coupling constant vanishes in the limit of large cut-off \cite{Fernandez:1992jh}. This is a particular case of the so-called quantum triviality:
all four-dimensional scalar theories are believed to be free. Although the situation is more subtle in the case in which the Higgs is coupled to other particles,
the question of triviality of the purely scalar sector of the SM still holds once one considers the perturbative expansion in the
gauge coupling constants keeping $\lambda$ fixed. This is known as the Higgs triviality problem.
On the other hand, since the Higgs mass $m$ is $125$ GeV, and $\langle\phi\rangle=m/(\sqrt{2\lambda})\simeq$ 246 GeV, it follows that according to
the usual formulation of the SM one should have for the scalar self-coupling
\be
\lambda\simeq 0.13 \ .
\ee
It should be stressed that the check of self-interactions terms in the Higgs legrangian are a priority in the LHC experiments.  This is a hot topic, for example
in \cite{Degrassi:2016wml} it has been proposed that the Higgs trilinear self-coupling can be tested in a near future via single Higgs production at LHC .

\vspace{0.2cm}

\noindent A natural question, which is also of considerable experimental interest, and suggested by the above analysis,
is whether it is possible to have a Higgs mechanism from a scalar model where the unique self-interaction is represented by the mass term.
Presumably such a possibility has not been considered because of
the non-negative definiteness of the Higgs field $\phi$ and the related assumption that the potential should be a function of
$\Phi^\dag\Phi\in\RR_{\geq0}$, with $\Phi$ the complex scalar doublet $\Phi$. We will see that these questions can be solved.

\vspace{0.2cm}

\noindent
We begin the investigation with the $U(1)$ Higgs-Kibble model, showing that one may generate the mass term with a free potential
and without SSB. In this respect, we note that an ingenious formulation of the Higgs phenomenon without SSB has been proposed by Fr\"ohlich, Morchio and
Strocchi in \cite{Frohlich:1980gj}\cite{Frohlich:1981yi}.
Then we will extend the analysis to the SM, formulating a mass generation without SSB. We will start with the analysis of the so-called polar decomposition of $\Phi$
\be
\Bigg(\begin{array}{c}\phi^+\\ \phi^0
\end{array}\Bigg) =U\Bigg(\begin{array}{c}0\\ \dfrac{\phi}{\sqrt2}\end{array}\Bigg) \ , \qquad \phi\in\RR_{\geq0} \ , \qquad U\in SU(2) \ .
\label{polar}\ee
Note that this parametrization does not imply any gauge choice and should not be confused with the unitary gauge. Also note
that $\phi=\sqrt{2\Phi^\dagger\Phi}$ is gauge invariant.

\vspace{0.2cm}

\noindent A key point related to the parametrization (\ref{polar}) is that the normal subgroup of $G=SU(3)\times SU(2)\times U(1)$, generated by
$(e^{2\pi i/3}I,-I,e^{\pi i/3})\in G$,
acts trivially on all the fields of the SM \cite{Hucks:1990nw,Baez:2005yf}. In particular, recalling that the action of the $U(1)$ is $e^{i3Y\gamma}$, and that $Y(\Phi)=1$,
one sees that in the case of $\Phi$, on which $SU(3)$ acts trivially, the $U(1)$ transformation with $\gamma=\pi/3$ times $-I\in SU(2)$ is the identity.
This suggests considering the representation (\ref{polar}) with $U(x)\in SU(2)/\ZZ_2\simeq SO(3)$,
where $\ZZ_2$ is the group generated by $-I$, and with $\phi\in\RR$ rather than
$\phi\in\RR_{\geq0}$. This solves the problem of the non-negative definiteness of $\phi$ of the polar decomposition.
It follows that while in the standard polar decomposition $\phi=\sqrt{2\Phi^\dag\Phi}$, in the case of $\RR\ni \phi\neq\sqrt{2\Phi^\dag\Phi}\in\RR_{\geq 0}$,
there are potentials\footnote{In the following the ``potential'' $U(\phi)$ denotes the mass term together with the true potential part $V(\phi)$.}
$U(\phi)$ not equivalent to $U(\sqrt{2\Phi^\dag\Phi})$.
In particular, one may consider potentials not constrained by the parity condition $U(-\phi)=U(\phi)$ and with $\phi\in\RR$.

\vspace{0.2cm}

\noindent
The outcome is that the simplest model for the Higgs mechanism, free
of the above mentioned problems of the standard formulation, is the one whose lagrangian density has the standard kinetic term
in addition to the mass term, with the right sign, and to a linear term in $\phi$. As a byproduct, we will also show that
the investigation provides the way to avoid Elitzur theorem in the usual formulation of the SM.

\vspace{0.2cm}

\noindent
In the remanent part of the paper we investigate a related model that still considers $\phi\in \RR$ and concerns the exponential interactions.
Recently, in \cite{Matone:2015nxa}, the exponential interaction has been considered as a master model to derive other scalar theories.
Interestingly, exponentiation of the Higgs also arises in the framework of skyrmions \cite{Atiyah}.

\vspace{0.2cm}

\noindent
The motivation for studying exponential interactions is that in the four-dimensional case, in agreement with the Higgs triviality problem, such theories
turns out to be free. In this sense such models are related to the above proposed model. Nevertheless, the precise correspondence between
such free models is still unknown, for this reason, considered the nice properties of the exponential interactions, it makes sense to investigate
their possible role.

\vspace{0.2cm}

\noindent
Let us recall that scalar theories with exponential interactions are non-renormalizable.
As emphasized in \cite{Pohlmeyer:1975gt}, the difficulties in quantizing some non-renormalizable field theories, concern the
non-uniqueness of the solution, rather than its existence.
In such a context, let us remind that in \cite{Albeverio:1979nc}, using the ultraviolet cut-off $\gamma^{-N}$, $\gamma>1$, $N>0$,
have been investigated scalar theories with interaction $\lambda :\exp(\alpha \phi):$. It turns out that for $D>2$, for all $\alpha$, and for $D=2$, with $|\alpha|>\alpha_0$, the Schwinger functions
converge to the free Schwinger functions. The essential point in the investigation of \cite{Albeverio:1979nc} is that
$\Delta_{F,\Lambda}(0)$, with $\Delta_{F,\Lambda}(x)$ the Feynman propagator
with cut-off on the momenta, grows sufficiently fast to kill the fluctuations of $\phi$, so that  $:\exp(\alpha \phi):= \exp(-{\alpha^2\over2}\Delta_{F,\Lambda}(0))\exp(\alpha \phi)$ vanishes in the limit
$\Lambda\to\infty$.

\vspace{0.2cm}

\noindent
In the present paper, we consider the $D$-dimensional euclidean scalar theory with potential
\be
V=\mu^D\exp(-\alpha\phi(x)) \ .
\ee
It turns out that the functional generator associated to such a potential is the first term of an expansion $W[J]=W_R[J]+\ldots$, where
\be
 W_R[J] =e^{-Z_R[J]}=\langle 0| :e^{-\int d^Dx V(\phi)}:|0\rangle_J \ ,
 \label{laf}\ee
and one may easily check that \cite{Matone:2015nxa}
 \be
Z_R[J]=Z_0[J]+\mu^D\int d^D x e^{-{\alpha}\int d^D y J(y)\Delta(y-x)} \ ,
\ee
where $\Delta(x-y)$ is the Feynman propagator.
It turns out that $Z_R[J]$ generates the lowest order contributions in $\alpha$ to the $N$-point point function. In particular
\be
-{\delta Z_R[J]\over\delta J(x)}|_{J=0}={\alpha\mu^D\over m^2} \ .
\ee
Next, as done in the previous model, we will parameterize the scalar doublet with $U(x)\in SU(2)/\ZZ_2$ and  $\phi\in\RR$ and consider the lagrangian density
\be
{\cal L}_\Phi=(D_\mu \Phi)^\dag (D^\mu \Phi)-{1\over2}m^2\phi^2+2\nu m^3 \sinh\Big({\phi\over\nu}\Big) \ .
\ee
We will see that this leads to
\be
\langle\phi\rangle=2m+{\cal O}(\nu^{-1}) \ .
\ee

\section{The $U(1)$ Higgs-Kibble model and Elitzur's theorem}

Let us consider the lagrangian density of the $U(1)$ Higgs-Kibble model
\be
{\cal L}=-{1\over 4} F_{\mu\nu} F^{\mu\nu}+{1\over2}(D_\mu\varphi)^\dagger(D^\mu\varphi)-U(|\varphi|) \ ,
\label{HiggsKibble}\ee
where $\varphi$ is a complex scalar and $D_\mu=\partial_\mu-ieA_\mu$. The fact that the potential depends only on $\rho=|\varphi|$ naturally selects the two independent
fields, $\rho$ and $\theta$, where $e^{i\theta}=\varphi/\rho$. So that (\ref{HiggsKibble}) is identical to
\be
{\cal L}=-{1\over 4} F_{\mu\nu} F^{\mu\nu}+{1\over2}e^2\rho^2 W_\mu W^\mu+{1\over2}\partial_\mu\rho\partial^\mu\rho-U(\rho) \ ,
\label{HiggsKibbleidenticadue}\ee
where $W_\mu=A_\mu+e^{-1}\partial_\mu\theta$. Note that a gauge transformation corresponds to $\theta\to \theta+\alpha$ and $A_\mu\to A_\mu-e^{-1}\partial_\mu\alpha$,
so that, like $\rho$, even $W_\mu$ is gauge invariant. In this way, without performing any gauge choice, one passes from the degrees of freedom $\varphi$ and $A_\mu$, to
$\rho$ and $W_\mu$.

\noindent
The usual treatment of (\ref{HiggsKibbleidenticadue}) is to consider a semiclassical approximation around the minimum $\rho_0$ of
$U(\rho)$. Set $\chi=\rho-\rho_0$. In such an approximation, considering only the terms quadratic in $\chi$ and $W_\mu$, one gets the lagrangian density
\be
\tilde{\cal L}=-{1\over 4} F_{\mu\nu} F^{\mu\nu}+{1\over2}e^2\rho_0^2 W_\mu W^\mu+{1\over2}\partial_\mu\chi\partial^\mu\chi-{1\over2}U''(\rho_0)\chi^2 \ ,
\label{resulting}\ee
showing that $W_\mu$ and $\chi$ have square masses $e^2\rho_0^2$ and $U''(\rho_0)$ respectively.

\noindent In the lucid analysis in \cite{Strocchi:2008gsa} have been discussed the main problems
with such a model. The first point is that in passing from (\ref{HiggsKibble}) to (\ref{resulting})
one has to fix the condition $\rho\in\RR_{\geq0}$, a difficult task even at the classic level because
this should be consistent with the time evolution. The problem is even more difficult in considering the semi-classical approximation because one should keep
$\chi$ bounded by $\rho_0$. At the quantum level there is the problem of treating the term $|\varphi|$. In a rigorous QFT formulation $\varphi$ is
a distribution and the modulus of a distribution is a ill-defined quantity. This implies that $\rho$ cannot be considered a quantum field.

\noindent An alternative approach is to make the decomposition $\varphi=\varphi_1+i\varphi_2$, and then considering the semi-classical expansion
\be
\varphi_1=\varphi_0+\chi_1 \ , \qquad \varphi_2= \chi_2 \ ,
\ee
with $\chi_1$ and $\chi_2$ considered as small fluctuations. The resulting lagrangian density is still (\ref{resulting}) with $W_\mu=A_\mu+e^{-1}\partial_\mu\chi_2$
and $\rho_0$ and $\chi$ replaced by $\varphi_0$ and $\chi_1$ respectively.

\noindent The problem with such a formulation is that while perturbation theory leads to $\langle\varphi\rangle\neq0$, at the non-perturbative level one has, according to
Elitzur's theorem \cite{Elitzur:1975im},
\be
\langle\varphi\rangle=0 \ .
\ee

\noindent Another possibility is to map $\varphi$ to a real field $\varphi_r\in\RR$ by a gauge transformation. Nevertheless, it turns out
that there is a residual $\ZZ_2$ gauge symmetry that gives, even in this case, $\langle\varphi_r\rangle=0$ \cite{Strocchi:2008gsa}. However, there is a key point which leads
to a well-defined solution. Namely, note that such a $\ZZ_2$ symmetry is the consequence of the tacitely assumed invariance of the
potential under $\varphi\to-\varphi$. On the other hand, one may interpret $e^{i\theta}$ and $\rho$ as independent fields, so that with $\rho\in \RR$, and
 $\theta\in (-\pi,\pi]$. An interesting alternative is to consider $\varphi=e^{i\theta}\rho$ as unique complex scalar field,
so that $\rho\in\RR$ requires $\theta\in (-\pi/2,\pi/2]$.
In the next section we will see that, in the case of the SM, the latter possibility is also suggested by the presence of a normal subgroup
of the gauge group leaving the fields invariant.

\noindent We can then consider $\rho$ to take real values and choose the potential in (\ref{HiggsKibble}) to be
\be
U(\rho)={1\over 2}m^2\rho^2-m^2\rho_0\rho \ ,
\label{potmin}\ee
so that the lagrangian density now reads
\be
{\cal L}=-{1\over 4} F_{\mu\nu} F^{\mu\nu}+{1\over2}(D_\mu \varphi)^\dagger(D^\mu \varphi)-{1\over 2}m^2\rho^2+m^2\rho_0\rho \ ,
\label{HiggsKibblemigliore}\ee
where $\varphi=e^{i\theta}\rho$, $\rho\in\RR$, and $\rho_0$ is a real constant. The purely scalar sector is now a free theory,
so that $\langle\rho\rangle$ coincides with the value of $\rho$ that minimizes (\ref{potmin}). Therefore, we have
\be
\langle \rho \rangle = \rho_0 \ .
\ee
Setting $\eta=\rho-\rho_0$ leads to the lagrangian density with the mass term for the gauge field without any SSB. In particular,
since $\rho$ is gauge invariant, the Elitzur theorem is avoided simply because it concerns the vacuum expectation value of gauge non-invariant quantities.

\section{Trivial Higgs}

Let us now consider the Higgs mechanism in the SM. By (\ref{polar})
we have that even in this case $\phi$ takes non-negative values. This means that $\eta=\phi-v$, $v=\langle\phi\rangle$, is bounded by
$-v$, so that the path integral on $\eta$ should be
\be
\int_{\eta\geq - v} D\eta e^{i S} \ .
\ee
Nevertheless, the field $\eta$ is usually considered as taking all real values. The standard argument to justify $\eta\in \RR$ is that one is considering
small oscillations around a minimum of the potential. We saw that this is a subtle point for several reasons.
In the case of the SM the condition $\eta\geq \nu$ has effects on all the terms
of the  lagrangian density of the SM, the kinetic one, the mass and the $\eta^3$ terms, and the Yukawa interactions.
Furthermore, even in doing perturbation theory, one should replace the Feynman propagator by the one coming from the path integral on field configurations bounded by $-v$.

\vspace{.2cm}

\noindent
Since the physical fields in the SM are the ones identified once one considers the polar decomposition,
it is clear that one should understand if an why one can choose the Higgs field $\eta$ to
take real values.
We now show that one may in fact relax the condition $\eta\geq-v$, a result that leads to the free model.
To this end, let us first recall that the reason why in (\ref{polar}) one can choose $U\in SU(2)$, that is
\be
U={\sqrt2\over\phi}\left(\begin{array}{c}{\bar\phi}^0\\ -{\bar\phi}^+
\end{array}\begin{array}{cc} \phi^+ \\ \phi^0\end{array}\right) \ ,
\ee
is because the first column of $U$ in the polar decomposition is completely arbitrary. Such an arbitrariness implies that the action on
$\Phi$ of a $U(1)$ transformation can be always represented by a matrix with the same determinant of $U$. In other words,
$$
e^{i\beta} \left(\begin{array}{c}a\\ c
\end{array}\begin{array}{cc}b\\ d\end{array}\right) \left(\begin{array}{c}0\\ 1
\end{array}\right) = \left(\begin{array}{c}e^{-i\beta} a\\ e^{-i\beta} c
\end{array}\begin{array}{cc} e^{i\beta} b\\ e^{i\beta} d\end{array}\right) \left(\begin{array}{c}0\\ 1
\end{array}\right) \ . $$
It follows that any $U(1)$ transformation of $\Phi$ is
equivalent to a map from $U\in SU(2)$ to $SU(2)$. In turn, this implies that
the $SU(2)\times U(1)$ and $SU(2)$ orbits of $\Phi$ are the same. In other words, any $SU(2)\times U(1)$ gauge transformation of $\Phi$
corresponds to a map of $U\in SU(2)$ to $U'\in SU(2)$, that is the gauge transformations act on $U$ only.
However, one should note that the identity transformation $\Phi\to\Phi$ is obtained in two different ways, by the simultaneous action of
the $U(1)$ and the $SU(2)$ identities and by acting with $-1\in U(1)$ and $-I\in SU(2)$.
This is related to the fact that the order 6 normal subgroup $N$ of $SU(3)\times SU(2) \times U(1)$, generated by
\be
(e^{2\pi i/3}I,-I,e^{\pi i/3})\in SU(3)\times SU(2) \times U(1) \ ,
\ee
acts trivially on all the fields of the SM (recall that $U(1)=e^{i3Y\gamma}$ and $Y(\Phi)=1$).
Therefore, the non-trivial part of the gauge group of the SM is \cite{Hucks:1990nw,Baez:2005yf}
\be
SU(3)\times SU(2) \times U(1)/N \ .
\ee
This leads to consider $U(x)$ and $\phi\in\RR$ as fully independent degrees of freedom, with $\phi$ that, as in the polar decomposition,
is gauge invariant. The point is to use the following parametrization
\be
\Phi(x)=
U(x)\Bigg(\begin{array}{c}0\\ \dfrac{\phi(x)}{\sqrt2}
\end{array}\Bigg) \ , \qquad \phi(x)\in\RR\ , \qquad U(x)\in SU(2)/\ZZ_2\simeq SO(3) \ ,
\label{keypoint}\ee
so that now $\phi=\pm\sqrt{2(|\phi^\dag|^2+|\phi^0|^2)}$. This suggests a possible role of $\ZZ_2$ monopoles \cite{Tomboulis:1981kd}-\cite{Tong:2017oea}.
Note that the analogous representation for a complex number $z=x+iy$ is
\be
z=\chi e^{i\theta} \ , \qquad \chi\in \RR \ , \qquad \theta\in(-\pi/2,\pi/2] \ ,
\label{nat}\ee
that should be compared with the polar decomposition $z=\rho e^{i\alpha}$, $\rho\geq 0$.
The principal part of $\arg z=\alpha$,  denoted ${\rm Arg}\, z\in (-\pi,\pi]$, corresponds to $\arctan (y/x)$ for $x\geq 0$.
In the case $x<0$ one has
${\rm Arg}\, z=\arctan (y/x)+\pi$ if $y\geq0$ and ${\rm Arg}\, z=\arctan (y/x)-\pi$ if $y<0$. Comparison with (\ref{nat}) shows that the natural choice is to set $\theta=\arctan (y/x)+2k\pi$, $k\in\ZZ$,
that is
\be
z=\chi e^{i[\arctan (y/x)+2k\pi]} \ , \qquad \chi \in \RR \ .
\ee
Note that
\be
\chi=\rho e^{i[{\rm Arg}\, z-\arctan (y/x)+2k\pi]}=\pm \rho \ .
\ee

\vspace{.2cm}

\noindent
As a result, $\Phi$ factorizes in a unitary field $U(x)$
times the gauge invariant field $\phi(x)\neq\sqrt{2\Phi^\dagger \Phi}$, which now takes values in the full real axis, so that
it can be considered a quantum field.
In the following we show that this leads to a simple gauge invariant lagrangian providing a non-trivial vev $\langle\phi\rangle$.

\vspace{.2cm}

\noindent
In the usual formulation the aspects related to the non-negative definiteness arise in two contexts. The first one concerns the choice of the range of $\phi$ discussed above.
The other one is the tacit assumption that the potential should be a function of $\Phi^\dag\Phi$.
As done in the previous section we relax such a condition by considering the lagrangian density
\be
{\cal L}_\Phi=(D_\mu \Phi)^\dag (D^\mu \Phi)-U(\phi) \ ,
\label{scalars}\ee
without the constraint $U(-\phi)=U(\phi)$ that would be implied if one chooses $U(\sqrt{2\Phi^\dag \Phi})$.

\vspace{.2cm}

\noindent As in the case of the $U(1)$ model, the above analysis indicates that there is a natural candidate which is free of the problems associated to the
formulation of the Higgs mechanism.
Let us choose $U(\phi)={1\over2}m^2\phi^2-2m^3\phi$, so that the lagrangian density of the purely scalar sector is
\be
{\cal L}_\phi={1\over2}\partial_\mu\phi\partial^\mu\phi-{1\over2}m^2\phi^2+{2m^3\phi} \ ,
\label{quella}\ee
$\phi\in\RR$. A nice consequence is that the contributions to $\langle\phi\rangle$ from the purely scalar sector
can be evaluated exactly. Namely, since the theory is the free one, it follows that the vev $\langle\phi\rangle$, evaluated
with respect to (\ref{quella}),
coincides with the value of $\phi$ that minimizes $\phi^2-4m\phi$, that is
\be
\langle\phi\rangle = {2m} \ .
\label{daduem}\ee
Note that this choice is in agrement with the experimental data $\langle\phi\rangle\approx 2m$, with the difference $2m-\langle\phi\rangle$ which may
fit the corrections, that we discuss below, due to the contributions to $\langle\phi\rangle$ coming from the other fields in the SM.
Setting $\eta=\phi-2m\in \RR$ the lagrangian density of the purely scalar part becomes
\be
{\cal L}_\eta = {1\over2}\partial_\mu\eta\partial^\mu\eta-{1\over2}m^2\eta^2 \ .
\label{quellafinala}\ee
The exact value of $\langle\phi\rangle$ can be evaluated by first
considering the path integration on $\phi$ taking into account that the full contributions to the quadratic and linear terms in $\phi$ include fermions, gauge bosons and the remanent
bosonic fields in $\Phi$.
Denoting by $F_1 \phi$ and $F_2\phi^2/2$ the contributions of such fields, the complete lagrangian density for $\phi$ has the form
\be
{1\over2}\partial_\mu\phi\partial^\mu\phi+{1\over2}(F_2-m^2)\phi^2+(F_1+2m^3)\phi  \ ,
\ee
giving the classical equation of motion
\be
(\partial_\mu\partial^\mu+m^2-F_2(x))\phi(x)=F_1(x)+2m^3 \ .
\ee
It follows that the vacuum expectation value, $\bar\phi$, of $\phi$, obtained by integrating only over $\phi$ is
\be
\bar\phi(x)=\int d^4y G(x,y)(F_1(y)+2m^3) \ ,
\ee
where $G(x,y)$ is the Green function for the operator $(\partial_\mu\partial^\mu+m^2-F_2(x))$. Note that neglecting $F_1$ and $F_2$, $G(x,y)$ reduces to
$-\Delta(y-x)$, where
$\Delta(y-x)$ is the minkowskian Feynman propagator. Using $\int d^4y \Delta(y-x)=-1/m^2$, one may check that in this case $\bar\phi(x)$ reproduces (\ref{daduem}).
The exact value of $\langle\phi\rangle$ is then given by
\be
\langle \phi(x)\rangle=\int d^4y\langle  G(x,y)(F_1(y)+2m^3) \rangle_{\chi} \ ,
\ee
where the subscript $\chi$ denotes the path integration over the remanent fields.

\noindent
A byproduct of the previous analysis is that the parametrization with $\phi$ taking all real values can be extended also to the
usual formulation of the SM. In particular, even if $\langle\phi\rangle\neq0$, there is no
contradiction with the Elitzur theorem because $\phi$ is now a genuine quantum field and gauge invariant, so that there is no
SSB. This arises only at the perturbative level by introducing the gauge fixing term. Therefore, the Higgs lagrangian density of the SM can be expressed in the form
\be
{\cal L}_\Phi=(D_\mu \Phi)^\dag (D^\mu \Phi)+{1\over2}\mu^2 \phi^2-{\lambda\over4} \phi^4 \ ,
\label{scalarsSMUSUAL}\ee
with $U\in SU(2)/\ZZ_2$ and $\phi\in\RR$.

\section{Exponential interactions}

In the following we investigate, in the euclidean space, a model that considers again $\phi\in \RR$ and concerns the exponential interactions.
The motivation for such an analysis is that such theories
are free for $D>2$ \cite{Albeverio:1979nc}, so that they are related to the above proposed model.

\vspace{0.2cm}

\noindent Let us shortly review the investigation in \cite{Matone:2015nxa}. The notation follows the one in \cite{Ramond:1981pw}.
Define
\begin{equation}
\langle f(x_1,\ldots,x_n)\rangle_{x_j\ldots x_k}\equiv\int d^D x_j\ldots d^D x_k
f(x_1,\ldots,x_n) \ ,
\end{equation}
and denote by $\langle f(x_1,\ldots,x_n)\rangle$ integration of $f$ over $x_1,\ldots,x_n$.
Let
\begin{equation}
{}\Delta(x-y)=\int {d^Dp \over (2\pi)^D} {e^{ip(x-y)}\over p^2+m^2} \ ,
\end{equation}
be the Feynman propagator and set
\be
Z_0[J]=-{1\over2}\langle J(x)\Delta(x-y) J(y)\rangle \ .
\ee
To compute $W[J]$ we use Schwinger's method
\be
{} W[J]=N e^{-\langle V({\delta\over\delta J})\rangle}e^{-Z_0[J]} \ .
\label{diventa}\ee
The first step in \cite{Matone:2015nxa} has been the observation
that exponential interactions can be obtained by acting on $\exp(-Z_0[J])$ with power series in the operator
$\langle\exp (-\alpha \delta_J)\rangle$ whose action corresponds to a translation of $J$.
Consider the potential investigated in \cite{Matone:2015nxa}
with the opposite sign of $\alpha$
\be
V(\phi)=\mu^D e^{-\alpha\phi} \ .
\ee
The corresponding generating functional (we drop the constant $N$) is
\begin{align}
W[J] & = \exp\Big[-\mu^D\langle \exp (-\alpha {\delta\over \delta J})\rangle \Big] \exp(-Z_0[J])\cr
&=\sum_{k=0}^\infty {(- \mu^{D})^k\over k!}\langle \exp(-\alpha {\delta\over\delta J})\rangle^k \exp(-Z_0[J]) \ .
\label{Winiziale}\end{align}
By \cite{Matone:2015nxa}
\be
\exp(-\alpha{\delta\over \delta J(x)})\exp(-Z_0[J])=\exp(-Z_0[J-\alpha_x])\exp(-\alpha{\delta\over \delta J(x)})=\exp(-Z_0[J-\alpha_x]) \ ,
\label{doposedici}\ee
where
\begin{align}
Z_0[J-\alpha_x]&= -{1\over 2}\int d^D y\int d^D z (J(y)-\alpha\delta(x-y))\Delta(y-z)(J(z)-\alpha\delta(x-z))\cr
&=Z_0[J]-{\alpha^2\over2}\Delta(0)+\alpha \int d^Dy J(y)\Delta(y-x) \ ,
\label{zetazero}\end{align}
one gets
\begin{align}
&W[J]=\exp(-Z_0[J])\sum_{k=0}^\infty\Big[ {(-\mu^{D})^k\over k!}\exp\big({k\alpha^2\over2}\Delta(0)\big)\cr
&\int d^D z_1\ldots\int d^D z_k \exp \Big(-\alpha\int d^Dz J(z)\sum_{j=1}^k \Delta(z-z_j)+\alpha^2\sum_{j>l}^k\Delta(z_j-z_l)\Big)\Big] \ .
\label{equation}\end{align}
Let us show that the Feynman propagators appearing in this expression are related to normal ordering. Let us focus on $\exp\big({k\alpha^2\over2}\Delta(0)\big)$ and $\exp(\alpha^2\sum_{j>l}^k\Delta(z_j-z_l)\big)$.
In this respect, note that (\ref{Winiziale}) corresponds to the expansion of
$\exp\Big(-\int d^DxV(\phi)\Big)$ in the time-ordered vev,
that is
\begin{align}
W[J]&=\langle 0| T e^{-\mu^D\int d^Dx\exp(-\alpha\phi(x))}|0\rangle_J \cr
&=\sum_{k=0}^\infty {(- \mu^{D})^k\over k!} \int d^Dx_1\ldots \int d^D x_k \langle0| T e^{-\alpha\phi(x_1)}\ldots e^{-\alpha\phi(x_k)}|0\rangle_{J} \ ,
\label{re}\end{align}
where the vacua are the ones of the free scalar theory coupled to the external source $J$.
Comparison with (\ref{equation}) fixes the expression of
$\langle0| T e^{-\alpha\phi(x_1)}\ldots e^{-\alpha\phi(x_k)}|0\rangle_{J}$.
The fact that the normal ordering problem is the cause of some of the
infinities arising in perturbation theory, suggests considering
\begin{align}
W_R[J]&=\langle 0| :e^{-\mu^D\int d^Dx\exp(-\alpha\phi(x))}:|0\rangle_J \cr
&=\sum_{k=0}^\infty {(-\mu^{D})^k\over k!} \int d^Dx_1\ldots \int d^D x_k \langle0|:e^{-\alpha\phi(x_1)}\ldots e^{-\alpha\phi(x_k)}:|0\rangle_{J} \ .
\label{redue}\end{align}
Note that
$:e^{-\alpha \phi(x)}:= e^{-{\alpha^2\over 2}\Delta(0)}e^{-\alpha \phi(x)}$,
and
\be
T:e^{-\alpha \phi(x_1)}: \ldots :e^{-\alpha \phi(x_k)}:
=e^{\alpha^2 \sum_{j>l}^k\Delta(x_j-x_l)}:e^{-\alpha \phi(x_1)}\ldots e^{-\alpha \phi(x_k)}: \ .
\label{eck}\ee
Therefore,
\be
:e^{-\alpha\phi(x_1)}\ldots e^{-\alpha\phi(x_k)}:
=e^{-{\alpha^2}\big({k\over 2}\Delta(0)+\sum_{j>l}^k \Delta(x_j-x_l)\big)} T e^{-\alpha\phi(x_1)}\ldots e^{-\alpha\phi(x_k)} \ .
\label{equesta}\ee
It follows that the expansion on the right hand side of (\ref{redue}) exponentiates.
Actually, (\ref{equation}), (\ref{re}), (\ref{redue}) and (\ref{equesta}) yield
\be
W_R[J]=\exp(-Z_R[J]) \ ,
\label{equationdue}\ee
where
\be
Z_R[J]=Z_0[J]+\mu^D\int d^D x e^{-{\alpha}\int d^D y J(y)\Delta(y-x)} \ .
\label{ilrisultatorinormalizzato}\ee
Interestingly, removing the term $\exp(\alpha^2\sum_{j>l}^k \Delta(x_j-x_l))$, coming from the normal ordering in (\ref{eck}), is equivalent to remove a term
$\langle\exp (-\alpha {\delta\over \delta J})\rangle$ in (\ref{Winiziale}). To show this, recall
that for any suitable function $F$, if $A$ and $B$ are operators, then $A^{-1}F(B)A=F(A^{-1}BA)$. Therefore,
\begin{align}
W[J] & = \exp\Big[-\mu^D\langle \exp (-\alpha {\delta\over \delta J})\rangle \Big] \exp(-Z_0[J]) \cr
&=\exp(-Z_0[J])\exp\Big[-\mu^D \exp(Z_0[J])\langle \exp (-\alpha {\delta\over \delta J})\rangle  \exp(-Z_0[J])\Big] \cr
&=\exp(-Z_0[J])\exp\Big[-\mu_0^D \langle\exp\big(-\alpha \langle J(y)\Delta(x-y)\rangle_y\big)\rangle_x\langle \exp (-\alpha {\delta\over \delta J})\rangle\Big] \ ,
\label{Winizialetre}\end{align}
where in the last equality we used (\ref{doposedici}) and (\ref{zetazero}), and
\be
\mu_0^D=\mu^D \exp\Big({\alpha^2\over 2}\Delta(0)\Big) \ .
\label{massa}\ee
Eq.(\ref{Winizialetre}) differs from $W_R[J]$ by the term $\langle\exp (-\alpha {\delta\over \delta J})\rangle$ in the last member, and by the relabeling of $\mu_0$.
The latter is equivalent to consider the normal ordering of $\exp(-\alpha\phi)$.
Therefore,
\be
W[J,:e^{-\alpha\phi}:]=W_R[J]+\ldots \ ,
\ee
where the dots denote the terms in (\ref{equation}) coming from the expansion
\be
\sum_{n=1}^\infty{\alpha^{2n}\over n!}\Big(\sum_{j>l}^k\Delta(z_j-z_l)\Big)^n \ .
\label{alphaenne}\ee

\noindent Consider the field
\be
\phi_{\rm cl}(x):=-{\delta Z_R[J]\over\delta J(x)} \ ,
\ee
and note that by (\ref{ilrisultatorinormalizzato})
\be
\phi_{\rm cl}(x)=\langle J(y)\Delta(x-y)\rangle_y+{\alpha\mu^D}\langle \Delta(y-x)\exp\big(-\alpha
\langle J(z) \Delta(y-z)\rangle_z\big)\rangle_y \ ,
\label{phicl}\ee
that satisfies the equation of motion
\be
(-\partial_\mu\partial_\mu+m^2)\phi_{cl}(x)=J(x)+{\alpha\mu^D}\exp(-\alpha\langle J(y)
\Delta(x-y)\rangle_y) \ .
\ee
By (\ref{phicl}) it follows that
$\Gamma_R[\phi_{\rm cl}]=Z_R[J]-\langle J(x) \phi_{\rm cl}(x)\rangle_x$, reads
\be
\Gamma_R[\phi_{\rm cl}]=Z_R[J]-\langle J(x)\Delta(x-y)J(y)\rangle_{xy}-{\alpha\mu^D}\langle J(x)\Delta(x-y)\exp(-\alpha
\langle J(z) \Delta(z-y)\rangle_z)\rangle_{xy}
\ee
Furthermore, at the first order in $\alpha$
\be
\langle0|\phi(x)|0\rangle = -{\delta Z_R[J]\over\delta J(x)}|_{J=0}={\alpha\mu^D\over m^2} \ ,
\label{ordinedue}\ee
where we used $\langle \Delta(x-y)\rangle_y={1/m^2}$.
It follows that the higher derivatives of $Z_R[J]$, evaluated at $J=0$, correspond, to the lowest order contribution in the $\alpha$ expansion, to the connected Green functions associated to
\be
\eta(x)=\phi(x)-{\alpha\mu^D\over m^2} \ ,
\label{eta}\ee
that is
\be
-{\delta^N Z_R[J]\over \delta J(x_1)\ldots \delta J(x_N)}|_{J=0}=\langle 0|T\eta(x_1)\ldots\eta(x_N)|0\rangle_c \ ,
\label{Npoint}\ee
and by (\ref{ilrisultatorinormalizzato}), for $N>1$,
\be
\langle 0|T\eta(x_1)\ldots\eta(x_N)|0\rangle_c
= \delta_{N2}\Delta(x_1-x_2)+\alpha^N\mu^D\int d^D y \Delta(y-x_1)\cdots\Delta(y-x_N) \ .
\ee
Note that higher order contributions in  $\alpha$ come from the expansion (\ref{alphaenne}).

\section{$\sinh(\phi/\nu)$}

The above model can be extended to more general interactions, such as
\be
V(\phi)=\sum_{k=1}^n\mu_k^D \exp(\alpha_k\phi) \ .
\label{pot}\ee
In order to find the explicit expression of $W_R[J]$ in the case of the potential (\ref{pot}), one first notes that
the exact generating functional
\be
W[J] = \exp(-Z[J]) = \Big[\prod_{k=1}^n \exp[-\mu_k^D\langle \exp (\alpha_k {\delta\over \delta J})\rangle]\Big] \exp(-Z_0[J]) \ ,
\label{n}\ee
and then uses (\ref{Winizialetre}) iteratively. The first step is
\begin{align}
& \exp\Big[-\mu_n^D\langle \exp (\alpha_n {\delta\over \delta J})\rangle \Big] \exp(-Z_0[J]) \cr
& =
\exp(-Z_0[J])\exp\Big[-\mu_{n0}^D \langle\exp\big(\alpha_n \langle J(y)\Delta(x-y)\rangle_y\big)\rangle_x\langle \exp (\alpha_n {\delta\over \delta J})\rangle\Big] \ .
\end{align}
Repeating this for the remaining $n-1$ terms in (\ref{n}), makes it clear that
\be
W_R[J]=\exp(-Z_R[J])=\langle 0|:e^{-\int d^D x \sum_{k=1}^n\mu_k^D \exp(\alpha_k\phi(x))}:|0\rangle_J  \ ,
\label{reotto}\ee
is obtained from $W[J]$ by removing, from the final expression, the term $\langle \exp \big(\sum_{k=1}^n\alpha_k {\delta\over \delta J}\big)\rangle$ on the right hand side,
and by
canceling the $\exp\big(\sum_{k=1}^n\alpha_k^2 \Delta(0)\big)$ term. Such a cancelation is equivalent to relabel each $\mu_{k0}$ by $\mu_k$. It follows that
\be
Z_R[J]=Z_0[J]+\int d^D x\sum_{k=1}^n\mu_k^D e^{{\alpha_k}\int d^D y J(y)\Delta(y-x)} \ .
\label{il}\ee
We note that taking the normal ordering of $\exp(-\int d^Dx V(\phi))$ may lead to well-defined $Z_R[J]$,
even in cases when $V(\phi)$ is unbounded below. A particularly interesting case is the four-dimensional potential
\be
V(\phi)=-2\nu m^3\sinh\Big({\phi\over\nu}\Big) \ .
\label{sinh}\ee
By (\ref{il}), we have
\be
Z_R[J]=Z_0[J]-2\nu m^3 \int d^4 x \sinh \Big( {\phi_c(x)\over \nu }\Big) \ ,
\label{HIGGSS}\ee
where
\be
\phi_c(x)=\int d^4 y J(y)\Delta(y-x) \ ,
\ee
that satisfies the free classical equation of motion in the presence of the external source $J$, is a key quantity in the dual representation of $W[J]$ recently introduced in \cite{Matone:2015aib}.
Repeating the analysis leading to (\ref{ordinedue}), at the zero order in $\nu^{-1}$, (\ref{HIGGSS}) yields
\be
\langle\phi\rangle=2m \ ,
\label{p}\ee
so that, at the same order,
\be
2^{-1/4}G_F^{-1/2}=2m \ ,
\ee
in agreement with the LHC data. Note that
\be
\lim_{\nu\to\infty}V(\phi)=-2m^3\phi \ ,
\label{mammamia}\ee
so that, in this limit, (\ref{p}) corresponds to the value of $\phi$ that minimizes $m^2\phi^2/2+V(\phi)$.

\noindent Making the expansion in powers of $\nu^{-1}$, one sees that the lowest order contribution to the $(2N+1)$-point functions is
generated by $Z_R[J]$, so that, at this order
\be
\langle 0|T\eta(x_1)\ldots\eta(x_{2N+1})|0\rangle
= 2^{2N+1} \nu^{-2N} m^{3}\int d^4 y \Delta(y-x_1)\cdots\Delta(y-x_{2N+1}) \ ,
\ee
where $\eta(x)=\phi(x)-\langle0| \phi(x)|0\rangle$.
In the case of the $2N$-point functions, $Z_R[J]$ contributes to the lowest-order of the two-point function only, so that it gives the free propagator.

\section{Conclusions and perspectives}

We proposed two models for mass generation for fermions and gauge bosons without SSB. One is based on a scalar sector without self-interactions and the other
with exponential interactions. In both cases there is no need to start with an imaginary mass term, so that even the initial lagrangian density is physically meaningful.

\vspace{.2cm}

\noindent
The first model has the mass term and a linear term in $\phi$.
The model is also suggested by the strong evidence that $\lambda\phi^4$ is a free theory,
that is, the renormalized coupling constant vanishes in the large cut-off limit. This is a particular case of the mentioned Higgs triviality problem.
As a byproduct, we saw how Elitzur theorem may be avoided in the usual formulation of the SM.

\vspace{.2cm}

\noindent
We also investigated the exponential interactions.
We then focused on the potential (\ref{sinh}).
In the limit $\nu\to\infty$ this explicitly corresponds to the free theory. At the next order, the theory
is described by $Z_R$, that, besides the propagator of the free theory,
generates only the lowest order contributions to the $(2N+1)$-point functions, $N>1$.
As such, for large $\nu$, the model is well-described by $Z_R[J]$, so that $W_R[J]$ can be seen as describing an effective theory.
Interestingly, exponentiation of the Higgs has been also considered in the framework of skyrmions \cite{Atiyah}.

\vspace{.2cm}

\noindent
An intriguing feature of the investigation is that the parametrization of the Higgs may be related
to string theory. The reason is that such a parametrization
is related to the normal subgroup of the SM that acts trivially on the fields, which in turn, is related to Calabi-Yau manifolds \cite{Baez:2005yf}.

\vspace{.2cm}

\noindent Another aspect of the formulation that should be investigated concerns the induced electroweak symmetry breaking model in \cite{Galloway:2013dma,Chang:2014ida,Contino:2017moj}.
In particular, the potential
in (\ref{quella}) is reminiscent of the effective Higgs potential obtained by integrating out the heavy mass eigenstate.

\vspace{.2cm}

\noindent Other investigations suggested by the proposed model concern the possible connection with $\ZZ_2$ monopoles, see for example
\cite{Tomboulis:1981kd}-\cite{Tong:2017oea}.

\vspace{.2cm}

\noindent We note that the absence of the cubic and quartic Higgs self-interactions can be tested experimentally. In particular, it is reasonable that, in a near future,
 LHC will give some evidence about the possible absence of the $\eta^3$ term in the lagrangian density. This can be checked in the production of two Higgs, with a virtual Higgs decaying
in two real Higgs. The process to be investigated at LHC is of course $p+p\to H+H$. Their absence would be a fundamental check of the present model for the Higgs mechanism. We also note that possible
precision tests may be suggested by the present model.

\vspace{.2cm}

\noindent
Let us conclude by mentioning that it would be interesting to investigate whether the free model we proposed
could be related to an effective theory in which the Higgs field is a fermionic condensate. Of course this
is suggested by the BCS theory of superconductivity that greatly motivated the original papers on the Higgs mechanism.
In fact there is a strong analogy with superconductivity, whose lower energy states can be described by the analogue of the Higgs field.
In this respect, it is interesting to note that the energy gap above the Fermi sphere is described by an exponential function that provides an example of
energy hierarchy scale (see the excellent book by Strocchi \cite{Strocchi:1985cf}
for  an  account on the BCS theory). A key feature of the superconductivity is that the condensation energy, that is the difference between the ground state energy in the superconducting state
and the conducting state, is of order
$10^{-7} - 10^{-8}$ eV per electron, much less than the other energy scales of the metal, which are of order of 1-10 eV.
In \cite{Morchio:1981hq} it has been used a mechanism reminiscent of the BCS theory to propose a non-perturbative mechanism explaining the problem of gauge hierarchies.
The analogy between the BCS theory and Higgs mechanism in the SM has been also stressed in the recent review by Peskin \cite{Peskin:2015kka}.

\section*{Acknowledgements}  It is a pleasure to thank Franco Strocchi for key comments on the Higgs model and the anonymous referee for bringing Ref.\cite{Galloway:2013dma} to my attention.
I also thank  Antonio Bassetto,  Giulio Bonelli, Tommaso Dorigo, Giuseppe Degrassi, Kurt Lechner, Pieralberto Marchetti, Paride Paradisi, Paolo Pasti, Javi Serra,
 Dima Sorokin, Mario Tonin, Roberto Volpato and Andrea Wulzer for interesting discussions.

\end{document}